\documentclass[english]{article}
\usepackage[T1]{fontenc}
\usepackage[latin9]{inputenc}
\usepackage{geometry}
\usepackage{amsmath}
\usepackage{setspace}
\usepackage{amssymb}
\makeatletter
\newcommand{\lyxaddress}[1]{
\par {\raggedright #1
\vspace{1.4em}
\noindent\par}
}

\makeatother

\usepackage{babel}

\begin{document}

\title{Quaternion-Octonion $SU$(3) Flavor Symmetry}

\author{Pushpa$^{(1)}$, P. S. Bisht$^{\left(1,2\right)},$ Tianjun Li$^{\left(2\right)}$
and O. P. S. Negi$^{\left(1\right)}$}

\maketitle

\lyxaddress{\begin{center}
$(1)$ Department of Physics,\\
 Kumaun University, S. S. J. Campus, \\
Almora-263601 (Uttarakhand) India
\par\end{center}}

\lyxaddress{\begin{center}
$(2)$ Institute of Theoretical Physics,\\
 Chinese Academy of Sciences,\\
 P. O. Box 2735, Beijing 100080, \\
P. R. China
\par\end{center}}

\lyxaddress{\begin{center}
Email - pushpakalauni60@yahoo.co.in\\
 ps\_bisht 123@rediffmail.com \\
tli@itp.ac.cn\\
ops\_negi@yahoo.co.in
\par\end{center}}
\begin{abstract}
Starting with the quaternionic formulation of isospin $SU(2)$ group,
we have derived the relations for different components of isospin
with quark states. Extending this formalism to the case of $SU(3)$
group we have considered the theory of octonion variables. Accordingly,
the octonion splitting of $SU(3)$ group have been reconsidered and
various commutation relations for $SU(3)$ group and its shift operators
are also derived and verified for different iso-spin multiplets i.e.
$I$,\ $U$ and $V$- spins. 


\textbf{PACS NO}: 11.30.Hv: Flavor symmetries; 12.10-Dm: Unified field
theories and models of strong and electroweak interactions;

.
\end{abstract}

\section{Introduction}

\ \ \ \ \ \ \ \ \ \ \ \ \ According to celebrated Hurwitz
theorem \cite{key-1} there exits four - division algebra consisting
of $\mathbb{R}$ (real numbers), $\mathbb{C}$ (complex numbers),
$\mathbb{H}$ (quaternions) and $\mathcal{O}$ (octonions). All four
algebras are alternative with antisymmetric associators. Quaternions
were very first example of hyper complex numbers introduced by Hamilton
\cite{key-2,key-3} and octonions by Caley \cite{key-4} and Graves
\cite{key-5} having the significant impacts on mathematics and physics.
Quaternions are also described in terms of Pauli spin - matrices for
the non - Abelian gauge theory in order to unify \cite{key-6} electromagnetism
and weak forces within the electroweak $SU(2)\times U(1)$ sector
of standard model.Yet another complex system (i.e, Octonion) \cite{key-7,key-8,key-9,key-10,key-11}
also plays an important role in understanding the physics beyond the
strong interaction between color degree of freedom of quarks and their
interaction. A detailed introduction on the various aspects and applications
of Exceptional, Jordan, Division, Clifford and non commutative as
well as non associative algebras related to octonions has recently
been discussed by Castro\cite{key-12} by extending the octonionic
geometry (gravity) developed earlier by Marques-Oliveira \cite{key-13,key-14}.
Recently we have also developed \cite{key-15} the quaternionic formulation
of Yang \textendash{} Mill\textquoteright{}s field equations and octonion
reformulation of quantum chromo dynamics $(QCD)$ where the resemblance
between quaternions and $SU(2)$ and that of octonions and $SU(3)$
gauge symmetries has been discussed. The color group $SU(3)_{C}$
is embedded with in the octonionic structure of the exceptional groups
while the $SU(3)$ flavor group has been discussed in terms of triality
property of the octonion algebra. Keeping in view, the utility of
quaternions and octonions, in this paper, we have made an attempt
to discuss the quaternionic formulation of isospin $SU(2)$ group
and octonion reformulation of $SU(3)$ flavor group (parallel to the
$SU(3)$ colour group \cite{key-15}). Accordingly, we have derived
the relations for different components of isospin with quark states.
Extending this formalism to the case of $SU(3)$ flavor group, the
octonion splitting of $SU(3)$ group have been reconsidered and various
commutation relations for $SU(3)$ group and its shift operators are
also derived and verified in terms of different iso-spin multiplets
i.e. $I$,\ $U$ and $V$- spins.

\section{Quaternionic representation of isospin $SU(2)$ group}

Let us consider the standard Lie algebra in terms of the quaternions
by adopting the basic formula of the isotopic formalism for nucleons.
Accordingly we may write a spinors $\psi_{a}$ and $\psi_{b}$ as\begin{eqnarray}
\psi & = & \left(\begin{array}{c}
\psi_{a}\\
\psi_{b}\end{array}\right);\label{eq:1}\end{eqnarray}
whereas the adjoint spinor is described as

\begin{eqnarray}
\overline{\psi} & = & \left(\begin{array}{cc}
\overline{\psi_{a}} & \overline{\psi}_{b}\end{array}\right);\label{eq:2}\end{eqnarray}
where 

\begin{align}
\psi=( & \psi_{0}+e_{1}\psi_{1})+e_{2}(\psi_{2}-e_{1}\psi_{3})=\psi_{a}+e_{2}\psi_{b}.\label{eq:3}\end{align}
In equations (\ref{eq:1}) and (\ref{eq:2}), $\psi_{a}=(\psi_{0}+e_{1}\psi_{1})$
and $\psi_{b}=(\psi_{2}-e_{1}\psi_{3})$ are considered over the field
of a quaternion described in terms of the four dimensional representations
of real numbers and two dimensional representations of complex numbers.
Accordingly, under $SU(2)$ gauge symmetry, the quaternion spinor
transforms as \cite{key-15,key-16},

\begin{eqnarray}
\psi & \longmapsto\psi^{\shortmid}= & U\,\psi\label{eq:4}\end{eqnarray}
where $U$ is $2\times2$ unitary matrix and satisfies 

\begin{eqnarray}
U^{\dagger}U & =UU^{\dagger}=UU^{-1}=U^{-1}U= & 1.\label{eq:5}\end{eqnarray}
On the other hand, the quaternion conjugate spinor transforms as \begin{eqnarray}
\overline{\psi}\longmapsto\overline{\psi^{\shortmid}} & = & \overline{\psi}U^{-1}.\label{eq:6}\end{eqnarray}
Hence, the combination $\psi\overline{\psi}=\overline{\psi}\psi=\psi\overline{\psi^{\shortmid}}=\overline{\psi^{\shortmid}}\psi$
is an invariant quantity. So, any unitary matrix may then be written
as 

\begin{eqnarray}
U & = & \exp\left(i\,\hat{H}\right);\label{eq:7}\end{eqnarray}
where $\widehat{H}$ is Hermitian $\widehat{H}^{\dagger}=\widehat{H}$.
Thus, we may express the Hermitian $2\times2$ matrix in terms of
four real numbers, $a_{1},\, a_{2},\, a_{3},$ and $\theta$ as 

\begin{eqnarray}
\hat{H} & = & \theta\hat{1}+\tau_{j}a_{j}=\theta1+ie_{j}a_{j};\label{eq:8}\end{eqnarray}
where $\hat{1}$ is the $2\times2$ unit matrix, $\tau_{j}$ are well
known $2\times2$ Pauli - spin matrices and $e_{1},\, e_{2},\, e_{3}$
are the quaternion units related to Pauli - spin matrices as \begin{eqnarray}
e_{0}= & 1;\,\,\,\,\, & e_{j}=-i\tau_{j}.\label{eq:9}\end{eqnarray}
Here $e_{j}$ $\left(\forall\, j=1,2,3\right)$, the basis elements
of quaternion algebra, satisfy the following multiplication rule

\begin{eqnarray}
e_{j}e_{k}=-\delta_{jk}+ & \epsilon_{jkl} & e_{l}\,\,(\forall j,k,l=1,2,3).\label{eq:10}\end{eqnarray}
where $\delta_{jk}$ and $\epsilon_{jkl}$ are respectively known
as usual Kronecker delta and three index Levi - Civita symbols respectively.
We may now describe the $SU(2)$ isospin in terms of quaternions as

\begin{align}
I_{a}= & \frac{ie_{a}}{2}\,\,\,(\forall\, a=1,2,3)\label{eq:11}\end{align}
and

\begin{align}
I_{\pm}= & \frac{i}{2}\left(e_{1}\pm ie_{2}\right).\label{eq:12}\end{align}
So, we may write the quaternion basis elements in terms of $SU(2)$
isospin as

\begin{align}
e_{1}=\frac{1}{i}\left(I_{+}+I_{-}\right);\,\,\,\,\,\,\, e_{2}=\frac{1}{i}\left(I_{+}-I_{-}\right);\,\,\,\,\,\,\, & e_{3}=\frac{1}{i}\left(I_{3}\right);\label{eq:13}\end{align}
which satisfy the following commutation relation

\begin{align}
\left[I_{+},I_{-}\right]=ie_{3};\,\,\,\quad & \left[I_{3},I_{\pm}\right]=\pm\frac{i}{2}\left(e_{1}\pm ie_{2}\right).\label{eq:14}\end{align}
 Here $SU(2)$ group acts upon the fundamental representation of $SU(2)$
doublets of up $\left(u\right)$ and down$\left(d\right)$ quark spinors 

\begin{align}
\left|u\right\rangle =\left(\begin{array}{c}
1\\
0\end{array}\right);\,\,\,\,\,\,\,\, & \left|d\right\rangle =\left(\begin{array}{c}
0\\
1\end{array}\right).\label{eq:15}\end{align}
So, we get for up quarks 

\begin{align}
I_{+}\left|u\right\rangle = & \frac{i\left(e_{1}+ie_{2}\right)}{2}\left|u\right\rangle =0;\,\,\,\,\,\,\, I_{-}\left|u\right\rangle =\frac{i\left(e_{1}-ie_{2}\right)}{2}\left|u\right\rangle =\frac{1}{2}\left|d\right\rangle ;\,\,\,\,\,\, I_{3}\left|u\right\rangle =\frac{ie_{3}}{2}\left|u\right\rangle =\frac{1}{2}u\left|u\right\rangle ;\label{eq:16}\end{align}
and for down quarks we have\begin{align}
I_{+}\left|d\right\rangle =\frac{i\left(e_{1}+ie_{2}\right)}{2}\left|d\right\rangle =\frac{1}{2}\left|u\right\rangle ;\,\,\,\,\, I_{-}\left|d\right\rangle =\frac{i\left(e_{1}-ie_{2}\right)}{2}\left|d\right\rangle =0;\,\,\,\,\,\,\, & I_{3}\left|d\right\rangle =\frac{ie_{3}}{2}\left|d\right\rangle =-\frac{1}{2}\left|d\right\rangle .\label{eq:17}\end{align}
Conjugates of equations$\left(\ref{eq:16}\right)$ and $\left(\ref{eq:17}\right)$
are now be described as

\begin{align}
\left\langle d\right|I_{+}=\left\langle d\right|\frac{i\left(e_{1}+ie_{2}\right)}{2}=0;\,\,\,\,\,\,\, & \left\langle d\right|I_{-}=\left\langle d\right|\frac{i\left(e_{1}-ie_{2}\right)}{2}=\frac{1}{2}\left\langle u\right|;\,\,\,\,\,\,\,\left\langle d\right|I=\left\langle d\right|\frac{ie_{3}}{2}=\frac{1}{2}\left\langle d\right|;\label{eq:18}\end{align}
and\begin{align}
\left\langle u\right|I_{+}=\left\langle u\right|\frac{i\left(e_{1}+ie_{2}\right)}{2}=\frac{1}{2}\left\langle d\right|;\,\,\,\,\,\,\left\langle u\right|I_{-}=\left\langle u\right|\frac{i\left(e_{1}-ie_{2}\right)}{2}=0;\,\,\,\:\,\,\,\, & \left\langle u\right|I_{3}=\left\langle u\right|\frac{ie_{3}}{2}=\frac{1}{2}\left\langle u\right|;\label{eq:19}\end{align}
where $e_{1}$, $e_{2}$ and $e_{3}$ are defined in $\left(\ref{eq:11}\right)$and
$\left(\ref{eq:12}\right).$ The effect of quaternion operator on
up $\left|u\right\rangle $ and down $\left|d\right\rangle $ quarks
states leads to

\begin{align}
ie_{1}\left|u\right\rangle =\left|d\right\rangle ;\,\,\,\,\,\,\, & ie_{1}\left|d\right\rangle =\left|u\right\rangle ;\nonumber \\
e_{2}\left|u\right\rangle =\left|d\right\rangle ;\,\,\,\,\,\,\, & e_{2}\left|d\right\rangle =-\left|u\right\rangle \nonumber \\
ie_{3}\left|u\right\rangle =\left|u\right\rangle ;\,\,\,\,\,\,\, & ie_{3}\left|d\right\rangle =-\left|d\right\rangle .\label{eq:20}\end{align}
So, we may write

\begin{align}
e_{1}\left(\begin{array}{c}
u\\
d\end{array}\right)= & i\left(\begin{array}{c}
u\\
d\end{array}\right);\label{eq:21}\end{align}

\begin{align}
e_{2}\left(\begin{array}{c}
u\\
d\end{array}\right)= & \left(\begin{array}{c}
d\\
-u\end{array}\right);\label{eq:22}\end{align}
\begin{align}
e_{3}\left(\begin{array}{c}
u\\
d\end{array}\right)= & i\left(\begin{array}{c}
u\\
-d\end{array}\right);\label{eq:23}\end{align}
transform a neutron (down quark) state into a proton ( up quark) state
or vice versa. Only $e_{2}$ gives real doublets of up and down quarks.

\section{Octonions and Gelmann $\lambda$ Matrices}

Octonions describe the widest normed algebra after the algebra of
real numbers, complex numbers and quaternions. The octonions are non
associative and non commutative normed division algebra over the algebra
of real numbers. A set of octets $(e_{0},\, e_{1},\, e_{2},\, e_{3},\, e_{4},\, e_{5},\, e_{6},\, e_{7})$
are known as the octonion basis elements and satisfy the following
multiplication rules

\begin{align}
e_{0}=1;\,\, e_{0}e_{A}=e_{A}e_{0}= & e_{A};\nonumber \\
e_{A}e_{B}=-\delta_{AB}e_{0}+f_{ABC}e_{C}.\,\,(\forall\, A,\, B,\, C=1,\,2,.....,\,7) & .\label{eq:24}\end{align}
The structure constants $f_{ABC}$ is completely antisymmetric and
takes the value $\mbox{\ensuremath{1}}$ for the following combinations,

\begin{eqnarray}
f_{ABC} & = & +1;\nonumber \\
\forall & (ABC) & =(123),\,(471),\,(257),\,(165),\,(624),\,(543),\,(736).\label{eq:25}\end{eqnarray}
The relation between Gell - Mann $\lambda$ matrices and octonion
units are given as \cite{key-15}

\begin{equation}
\frac{\left[e_{a+3},\, e_{7}\right]}{\left[\lambda_{a+3},\,\lambda_{7}\right]}=\frac{e_{a}}{2i\lambda_{a}};\label{eq:26}\end{equation}

\begin{equation}
\frac{\left[e_{7},\, e_{a}\right]}{\left[\lambda_{7},\,\lambda_{a}\right]}=\frac{e_{a+3}}{2i\lambda_{a+3}};\label{eq:27}\end{equation}

\begin{equation}
\frac{\left[e_{a},\, e_{a+3}\right]}{\left[\lambda_{a},\,\lambda_{a+3}\right]}=\frac{e_{7}}{2i\lambda_{7}};\label{eq:28}\end{equation}
where $a=1,\,2,\,3$. So, we have may describe \cite{key-15} the
following relationship between Gell - Mann $\lambda$ matrices and
octonion units as , 

\begin{align}
\lambda_{1}=-ie_{1}\, k_{1};\,\lambda_{2}=-ie_{2}\, k_{2};\,\lambda_{3}=-ie_{3}\, k_{3}; & \,\lambda_{4}=-ie_{4\,}k_{4};\nonumber \\
\lambda_{5}=-ie_{5\,}k_{5};\,\lambda_{6}=-ie_{6\,}k_{6}; & \,\lambda_{7}=-ie_{7}\, k_{7};\label{eq:29}\end{align}
and $\lambda_{8}$ are also related with $e_{3}$ as

\begin{equation}
\lambda_{8}=-ik_{8}e_{3}\,\,\, where\,\,\, k_{8}=\frac{8}{\sqrt{3}};\label{eq:30}\end{equation}
where $k_{a}=-1\,\,\,(\forall a=1,2,3,.....,7)$ are proportionality
constants.

\section{Octonion Formulation of $SU(3)$ Flavor Group }

The Lie algebra of $SU(3)$ exhibits most of the features of the larger
Lie algebras. The elements of $SU(3)$ group may be obtained in terms
of $3\times3$ Hermitian Gell Mann $\lambda$ Hermitian matrices related
to octonions where first three matrices describe the familiar isotopic
spin generators from the $SU(2)$ subgroup of $SU(3)$. These generators
connect up $(u)$ and down quarks $(d)$. The fourth and fifth generators
and the sixth and seventh generators are denoted as the $V$ - spin
and the $U$ - spin. $V$ - spin connects the up $(u)$ and strange
quarks $(s)$ while $U$ - spin connects the down $(d)$ and strange
quarks $(s)$. The eighth generator is diagonal in nature responsible
for hypercharge. The $I$, $U$ and $V-$ spin algebra fulfills the
angular momentum algebra and turn out to the sub algebras of $SU(3)$.
Hence, the $SU(3)$ multiplets are constructed in from of a $I-$
multiplets, $V-$ multiplets and an $U-$ multiplets. The $I-$ spin,
$U-$ spin and $V-$spin algebra are closely related and are the elements
of sub algebra of $SU(3)$. As we may define $SU(3)$ multiplets as

\begin{eqnarray}
I_{1}=\frac{1}{2}\lambda_{1}=\frac{ie_{1}}{2}; & I_{2}=\frac{1}{2}\lambda_{2}=\frac{ie_{2}}{2}; & I_{3}=\frac{1}{2}\lambda_{3}=\frac{ie_{3}}{2}\,(I-\, Spin)\nonumber \\
V_{1}=\frac{1}{2}\lambda_{4}=\frac{ie_{4}}{2}; & V_{2}=\frac{1}{2}\lambda_{5}=\frac{ie_{5}}{2}; & V_{3}=\frac{ie_{3}}{4}\left(8\sqrt{3}+1\right)\,(U-\, Spin)\nonumber \\
U_{1}=\frac{1}{2}\lambda_{6}=\frac{ie_{6}}{2}; & U_{2}=\frac{1}{2}\lambda_{7}=i\frac{e_{7}}{2}; & U_{3}=\frac{ie_{3}}{4}\left(8\sqrt{3}-1\right)\,(V-\, Spin)\label{eq:31}\end{eqnarray}
along with the hyper charge is written in terms of octonions as follows\begin{equation}
Y=\frac{1}{\sqrt{3}}\lambda_{8}=-\frac{8ie_{3}}{\sqrt{3}}.\label{eq:32}\end{equation}
Here $I_{1}$, $I_{2}$ and $I_{3}$ contain the $2\times2$ isospin
operators (i. e. quaternion units). $U_{3}$ ,$V_{3}$,$\, I_{3}$
and $Y$ are linearly independent generators and are simultaneously
diagonalized. It will to be noted that $\lambda_{1}$,\ $\lambda_{2}$,$\,\lambda_{3}$
agree with $\sigma_{1}$,$\,\sigma_{2},$$\,\sigma_{3}.$ The matrices
$V_{1}$,$\, V_{2}$,$\, U_{3}$ and $U_{2}$ connect the nucleon
and $\Lambda$ and changes strangeness by one unit and isotopic spin
by a half unit.. The operators $I_{\pm}$,$\, U_{\pm},\, V_{\pm}$
are also the shift operators known as raising ($I_{+}$,$\, U_{+}$,\ $V_{+})$
and lowering ($I_{-}$,$\, U_{-}$,$\, V_{-}$) operators. $I_{\pm}$,$\,\, U_{\pm}$,$\,\, V_{\pm}$
are called as ladder operators. The complexified variants contain
the third operators $T_{\pm}$,$\,\, U_{\pm}$,$\,\, V_{\pm}$ which
characterizes the states of $SU(3)$ multiplets.The operators $I_{\pm}$,\ $U_{\pm},\, V_{\pm}$
are defined as

\begin{align}
I_{\pm}=I_{x}\pm iI_{y}= & \frac{1}{2}\left(\lambda_{1}\pm i\lambda_{2}\right)=\frac{i}{2}\left(e_{1}\pm ie_{2}\right).\label{eq:33}\end{align}

\begin{align}
V_{\pm}=V_{x}\pm iV_{y}= & \frac{1}{2}\left(\lambda_{4}\pm i\lambda_{5}\right)=\frac{i}{2}\left(e_{4}\pm ie_{5}\right).\label{eq:34}\end{align}

\begin{align}
U_{\pm}=U_{x}\pm iU_{y}= & \frac{1}{2}\left(\lambda_{6}\pm i\lambda_{7}\right)=\frac{i}{2}\left(e_{6}\pm ie_{7}\right).\label{eq:35}\end{align}
With the help of shift operators and their properties ,we may derive
the quark states of these multiplets as $\left|q_{1}\right\rangle $$\left|q_{2}\right\rangle $$\left|q_{3}\right\rangle $.So
the quark states of $I$ ,$U$ and $V$ spin are described as\begin{align}
I_{-}\mid q_{1}>=\mid q_{2}>;\,\,\,\,\,\,\, & I_{+}\mid q_{2}>=\mid q_{1}>.\label{eq:36}\end{align}

\begin{align}
U_{-}\mid q_{2}>=\mid q_{3}>;\,\,\,\,\,\,\, & U_{+}\mid q_{3}>=\mid q_{2}>.\label{eq:37}\end{align}

\begin{align}
V_{-}\mid q_{1}>=\mid q_{3}>;\,\,\,\,\,\,\, & V_{+}\mid q_{3}>=\mid q_{1}>.\label{eq:38}\end{align}
Thus,the operators $I_{\pm}$,\ $U_{\pm},\, V_{\pm}$ are viewed
as operators which transforms one flavor into another flavor of quarks 

\begin{align}
I_{\pm}\left(I_{3}\right)\longmapsto & I_{3}\pm1;\nonumber \\
V_{\pm}\left(V_{3}\right)\longmapsto & V_{3}\pm1;\nonumber \\
U_{\pm}\left(U_{3}\right)\longmapsto & U_{3}\pm1.\label{eq:39}\end{align}
It means the action of $I_{\pm}$ ,$U_{\pm}$ and $V_{\pm}$ shifts
the values of $I_{3}$ ,$V_{3}$ and $U_{3}$ by $\pm1.$

\section{Commutation relations for Octonion Valued Shift Operators }

The $I$, $U$ and $V$ - spin algebras are closed. Let us obtain
the commutation relations of shift operators $I_{\pm},\, U_{\pm},\, V_{\pm}$
\cite{key-16} for $SU(3)$ group.Using equations (\ref{eq:24}),
(\ref{eq:25}),(\ref{eq:34}) and (\ref{eq:35}),we get

\begin{align}
\left[U_{3},\, U_{\pm}\right]=\pm\left[\frac{i}{2}\left(e_{6}\pm ie_{7}\right)\right]= & \pm U_{\pm};\nonumber \\
\left[V_{3},\, V_{\pm}\right]=\pm\left[\frac{i}{2}\left(e_{4}\pm ie_{5}\right)\right]= & \pm V_{\pm};\nonumber \\
\left[I_{3},\, I_{\pm}\right]=\pm\left[\frac{i}{2}\left(e_{1}\pm ie_{2}\right)\right]= & \pm I_{\pm}.\label{eq:40}\end{align}

\begin{align}
\left[U_{+},\, U_{-}\right]=\frac{ie_{3}}{2}\left(8\sqrt{3}-1\right) & =2U_{3};\nonumber \\
\left[V_{+},\, V_{-}\right]=\frac{ie_{3}}{2}\left(8\sqrt{3}+1\right) & =2V_{3};\nonumber \\
\left[I_{+},\, I_{-}\right]=ie_{3} & =2I_{3}.\label{eq:41}\end{align}

\begin{align}
\left[Y,\, U_{\pm}\right]=\pm\left[\frac{i}{2}\left(e_{6}\pm ie_{7}\right)\right]= & \pm U_{\pm};\nonumber \\
\left[Y,\, V_{\pm}\right]=\pm\left[\frac{i}{2}\left(e_{4}\pm ie_{5}\right)\right]= & \pm V_{\pm};\nonumber \\
\left[Y,\, I_{\pm}\right]=\pm\left[\frac{i}{2}\left(e_{1}\pm ie_{2}\right)\right]= & \pm I_{\pm}.\label{eq:42}\end{align}

\begin{align}
\left[I_{+},\, V_{+}\right]=\left[I_{+},U_{+}\right]=\left[U_{+},V_{+}\right]= & 0;\label{eq:43}\end{align}

\begin{align}
\left[I_{+},\, V_{-}\right]=-\left[\frac{i}{2}\left(e_{6}-ie_{7}\right)\right] & =-U_{-};\nonumber \\
\left[T_{+},U_{+}\right]=\left[\frac{i}{2}\left(e_{4}+ie_{5}\right)\right] & =V_{+};\nonumber \\
\left[U_{+},\, V_{-}\right]=\frac{i}{2}\left(e_{1}-ie_{2}\right) & =I_{-}.\label{eq:44}\end{align}
Accordingly, we may write the hypercharge as

\begin{align}
Y= & \frac{1}{\sqrt{3}}\lambda_{8}=-\frac{8ie_{3}}{\sqrt{3}}=\frac{2}{3}\left(U_{3}+V_{3}\right)=\frac{2}{3}\left(2U_{3}+I_{3}\right)=\frac{2}{3}\left(2V_{3}-I_{3}\right).\label{eq:45}\end{align}
and the term hyper charge $Y$ commutes with third component of $I$,
$U$ and $V$ - spin multiplets of $SU(3)$ flavor group

\begin{align}
\left[Y,\, I_{3}\right]=\left[Y,\, U_{3}\right]=\left[Y,\, V_{3}\right]= & 0\label{eq:46}\end{align}
Under the following dependency conditions,

\begin{align}
I_{3}= & \frac{ie_{3}}{2}=V_{3}-U_{3}\label{eq:47}\end{align}
and it also \begin{align}
I_{+}=\frac{i}{2}\left(e_{1}\pm ie_{2}\right) & =\left(I_{-}\right)^{\dagger}\nonumber \\
V_{+}=\frac{i}{2}\left(e_{4}\pm ie_{5}\right) & =\left(V_{-}\right)^{\dagger}\nonumber \\
U_{+}=\frac{i}{2}\left(e_{6}\pm ie_{7}\right) & =\left(U_{-}\right)^{\dagger}\label{eq:48}\end{align}
The commutation relation between the third components of $T$,$\, U$
and $V$ with $T_{\pm}$ are given as

\begin{align}
\left[I_{3},\, I_{\pm}\right]=\pm\left[\frac{i}{2}\left(e_{1}\pm ie_{2}\right)\right]= & \pm I_{\pm};\nonumber \\
\left[U_{3},\, I_{\pm}\right]=\mp\frac{1}{2}\left[\frac{i}{2}\left(e_{1}\pm ie_{2}\right)\right]= & \mp\frac{1}{2}I_{\pm};\nonumber \\
\left[V_{3},\, I_{\pm}\right]=\pm\frac{1}{2}\left[\frac{i}{2}\left(e_{1}\pm ie_{2}\right)\right] & =\pm\frac{1}{2}I_{\pm};\label{eq:49}\end{align}
The octonions are related to raising $I_{+}$,\ $U_{+}$,\ $V_{+}$
and lowering $I_{-}$,\ $U_{-}$,$\, V_{-}$ operators as

\begin{align}
e_{1}=i\left(I_{+}+I_{-}\right);\,\,\,\, e_{2}=\left(I_{+}-I_{-}\right); & \,\,\,\, e_{3}=2I_{3};\nonumber \\
e_{4}=i\left(V_{+}+V_{-}\right);\,\,\,\, & e_{5}=\left(V_{+}-V_{-}\right);\nonumber \\
e_{6}=i\left(U_{+}+U_{-}\right);\,\,\,\, & e_{7}=\left(I_{+}+I_{-}\right).\label{eq:50}\end{align}
The commutation relations between $I_{+}$ and $I_{-}$,\ $U_{+}$
and $U_{-}$and $V_{+}$ and $V_{-}$ are described as

\begin{align}
\left[I_{+},\, I_{-}\right]=ie_{3} & =2I_{3};\nonumber \\
\left[U_{+},\, U_{-}\right]=\frac{ie_{3}}{2}\left(8\sqrt{3}-1\right) & =2U_{3};\nonumber \\
\left[V_{+},\, V_{-}\right]=-\frac{ie_{3}}{4}\left(8\sqrt{3}+1\right) & =2V_{3}.\label{eq:51}\end{align}

\section{Discussion and Conclusion }

In the foregoing analysis, we have described the resemblance between
real Lie algebra $SU(2)$ and the quaternion formalism of isospin
$SU((2)$ group by using the isomorphism between quaternion basis
elements and Pauli matrices .On the same footing, we have established
the connection between octonion basis elements and Gell Mann $\lambda$
matrices of $SU(3)$ flavor group. Accordingly, we have discussed
the various sub algebras of $SU(3)$ flavor group constructed in terms
three $SU(2)$ algebras of $I$, $U$ and $V$ -spin multiplets. It
therefore appears that the $SU(3)$ flavor group may represent a new
fundamental symmetry using octonions for classifying hadrons within
$SU(3)$ multiplets.The isospin symmetries are good approximation
to simplify the interaction among hadrons. The motivation behind the
present theory was to develop a simple compact and consistent algebraic
formulation of $SU(2)$ and $SU(3)$ symmetries in terms of normed
algebras namely quaternions and octonions. So, we have described the
compact simplified notations instead of using the Pauli and Gellmann
matrices respectively got $SU(2)$ and $SU(3)$ groups. Octonion representation
of $SU(3)$ flavor group directly establishes the one to one mapping
between the non-associativity and the theory of strong interactions.
It also shows that the theory of hadrons (or quark-colour etc.) has
the direct link with non- associativity (octonions) while the isotopic
spin leads to non commutativity (quaternions). So, we have derived
and verified the commutation relations of shift operators and generators
of $SU(3)$flavour group using octonions. As such, normed algebras
namely the algebra of complex numbers, quaternions and octonions play
an important role for the physical interpretation of quantum electrodynamics
(QED), standard model of EW interactions and quantum chromo dynamics
(QCD). 

\textbf{ACKNOWLEDGMENT}: One of us PSB is thankful to Third World
Academy of Sciences for awarding him TWAS-UNESCO Associateship. He
is also thankful to Professor Yue-Liang Wu for his hospitality at
the Institute of Theoretical Physics, Beijing, China.

\end{document}